\documentstyle[12pt, aaspp4, flushrt]{article}

\begin{document}

\title{HST Observations of IP~Pegasi in Quiescence: The Pre-Eclipse Spectrum}

\author{D. W. Hoard,\footnotemark[1] 
Raymundo Baptista,\footnotemark[2] 
Michael Eracleous,\footnotemark[3]$^{,}$\footnotemark[8] 
Keith Horne,\footnotemark[4] 
K. A. Misselt,\footnotemark[5] 
Allen W. Shafter,\footnotemark[6] 
Paula Szkody,\footnotemark[1] 
Janet H. Wood\footnotemark[7]}
\footnotetext[1]{Department of Astronomy, University of Washington, Box 351580, 
Seattle, WA 98195-1580, USA}
\footnotetext[2]{Depto. F\'\i sica, UFSC, Campus Trindade, Florian\'{o}polis, 
88040-900, Brazil}
\footnotetext[3]{Department of Astronomy, University of California, Berkeley, 
CA 94720, USA}
\footnotetext[4]{School of Physics \& Astronomy, University of 
St.\ Andrews, KY16 9SS St.\ Andrews, Fife, Scotland}
\footnotetext[5]{Department of Physics and Astronomy, Louisiana State 
University, Baton Rouge, LA 70803-4001, USA}
\footnotetext[6]{Department of Astronomy, San Diego State University, 
San Diego, CA 92182, USA}
\footnotetext[7]{Department of Physics, Keele University, Keele, 
Staffordshire, ST5 5BG, UK}
\footnotetext[8]{Hubble Fellow}

\slugcomment{to appear in Monthly Notices of the Royal Astronomical Society}

\begin{abstract}

We present time-resolved HST ultraviolet spectroscopy and ground-based optical 
photometry of the dwarf nova IP Pegasi in a quiescent state.  The observations 
were obtained prior to an eclipse, when the bright spot caused by the impact of 
the accretion stream with the edge of the disk dominates the light output.  The 
optical light curve is fairly strongly correlated with the UV 
spectrophotometric flux curve.  An unusual emission-like feature near 1820 \AA\ 
in the UV spectrum of IP Peg is likely to be a manifestation of the 
``\ion{Fe}{2} curtain.''  Composite spectra constructed from the peaks and 
troughs of flickers in the light curve show substantial differences.  The 
spectrum of the flickers (i.e., peaks minus troughs) is not adequately modelled 
by a simple blackbody, suggesting that a more sophisticated model is 
appropriate.  We perform a cross-correlation analysis of the variability in 
spectrophotometric flux curves of the UV continuum and prominent UV emission 
lines (\ion{C}{2} $\lambda1335$, \ion{Si}{4} $\lambda1400$, \ion{C}{4} 
$\lambda1550$).  The continuum and lines are not correlated, suggesting that 
they are produced separately.  The \ion{C}{2} and \ion{Si}{4} lines are 
moderately correlated with each other, but neither line is correlated with 
\ion{C}{4}, suggesting that the latter forms in a different region than the 
former.  We briefly discuss a qualitative model for the geometry of the 
emission regions in IP Peg that is consistent with the observed behavior of the 
UV lines and continuum.

\end{abstract}


\section{Introduction}
\label{s-intro}
The cataclysmic variables (CVs) are semi-detached close binaries 
($P_{orb}\lesssim1$ d), in which mass is transferred from the 
Roche-lobe-filling, late main sequence secondary star into a rotating accretion 
disk around the white dwarf (WD) primary star.  Mass transfer from the inner 
Lagrangian point at the tip of the secondary star's Roche lobe to the outer 
edge of the disk occurs via an accretion stream that follows a nearly ballistic 
trajectory.  The dwarf novae are a subclass of CV that undergo outbursts with 
durations of a few days to a few weeks and repeat on time scales of a few tens 
to a few hundreds of days.  These outbursts are characterized by a rapid 
increase in brightness of 2--6 mag above the quiescent level.  The cause of the 
outbursts is thought to be related to one or more instabilities in the mass 
transfer and/or accretion processes (see the review by \markcite{Osaki96}Osaki 
1996).

A common feature of the CV class as a whole is the presence of random, short 
time scale ``flickering'' in their light output that is thought to be produced 
in the accretion disk+stream.  An additional (periodic) photometric modulation 
is often created by a bright spot that forms at the impact site of the 
accretion stream with the outer edge of the disk.  In high inclination CVs, this bright spot is typically located directly along our line-of-sight to the system 
shortly before the eclipse of the hot disk+WD by the cool secondary star. The 
resultant increase in the level of the system's light curve is often referred 
to as the ``pre-eclipse hump.''  The bright spot is also a probable source of 
flickering in CVs.

IP~Peg is an eclipsing dwarf nova with orbital period $3.8$~h.  The eclipse is 
very deep ($\Delta B\approx2.5$ mag) and the system shows a prominent 
pre-eclipse hump that begins near orbital phase $\phi=0.65$ and peaks at 
$\phi=0.8$--$0.9$ with $\Delta B\approx1$ mag (\markcite{Gor85}Goranskij, 
Orlowsky, \& Rahimov 1985).  \markcite{Marsh88}Marsh (1988) conducted a 
spectroscopic study of IP~Peg in the optical regime, and calculated the 
following system parameters from optical emission line radial velocities: 
$M_{WD}=1.09\pm0.10M_{\odot}$, $M_{2}=0.64\pm0.09M_{\odot}$, and 
$i=79.3\pm0.8^{\circ}$.  Marsh also isolated the spectrum of the bright spot by 
subtracting the average spectrum during phases outside both the pre-eclipse 
hump and the eclipse from a spectrum obtained at the peak of the hump.  A 
blackbody with temperature $T=12,000\pm1000$ K produces the best fit to the 
continuum of the bright spot spectrum.  Szkody (1987), on the other hand, 
used a similar method applied to an {\em International Ultraviolet Explorer} 
(IUE) spectrum of IP~Peg to determine a blackbody temperature of 
$T\approx20,000$ K for the bright spot.  Doppler tomography of IP~Peg in the 
Balmer lines shows a fairly uniform disk structure with a bright spot at the 
expected site of the accretion stream impact; the brightness of the spot varies 
in tomograms constructed from spectra obtained several years apart 
(\markcite{Marsh90}Marsh \& Horne 1990, \markcite{Kait94}Kaitchuck et al. 
1994).  This apparent variability of the bright spot radiation may explain the 
different temperatures determined by Szkody and Marsh, or it may be due to 
observing a temperature-stratified spot in different wavelength regimes -- the 
dwarf nova Z~Chamaeleontis showed a similar discrepancy between bright spot 
temperatures determined from optical ($T\approx11,300$ K; \markcite{Wood86}Wood 
et al.\ 1986) and UV ($T\approx16,700$ K; \markcite{Rob95}Robinson et al.\ 
1995) observations.

IP Peg was observed with the Faint Object Spectrograph (FOS) 
on the {\em Hubble Space Telescope} (HST)  
on 9 occasions between 1992 November and 1994 October. The first 7
observations are presented in Baptista et al.\ (1994), while a paper presenting 
the entire set of observations in detail is in preparation. Here we 
concentrate on the first observation, which was carried out on 1992 November 2 
UT.  The goal of this observation was to obtain time-resolved spectrophotometry 
during eclipse while the system was in a quiescent state.  Due to a scheduling 
error, however, Run~1 was obtained during ingress to eclipse rather than during 
the eclipse itself.  These pre-eclipse measurements can be utilized to study 
the rapid timescale flickering associated with the bright spot and beginning of 
eclipse in IP~Peg.  We report here on the results of our analysis of the Run~1 
observations and simultaneous ground-based photometry of IP~Peg.

\section{Observations} 
\subsection{HST Ultraviolet Spectra}
\label{s-hst_obs}
The observation started at UT 08:17 (HJD 2448928.84955) and spanned a single
spacecraft orbit. During the part of the orbit when the target was visible
to the telescope (40 minutes), 437 spectra were taken in rapid succession, with
an exposure time of 1.18~s and a time resolution of 5.5~s. 
The spectra were obtained through a $4 \farcs 3$ aperture, using the G160L 
grating, which gave a spectral resolution of 9.2 \AA\ (FWHM) over the range 
$1150-2500$ \AA. The blue ends of the spectra are severely contaminated by 
geocoronal Ly$\alpha$ emission, while the 2nd order geocoronal Ly$\alpha$
emission contaminates the red ends at the few percent level. Using the IP Peg 
eclipse ephemeris of \markcite{Wolf93}Wolf et al. (1993)
\begin{equation}
HJD_{min} = 2445615.4224(4) + 0.15820616(4)E,
\end{equation}
the observations cover the orbital phase range $\phi\approx 0.73$--$0.91$.

The data reduction consisted of background subtraction, flat field division,
and flux calibration. The background subtraction involved the scaling of the
nominal background spectrum to the observed time-dependent background level,
as recorded by the ``unilluminated'' segments of the detector array. Thus, 
this method corrects not only for the particle background but also for 
contamination by light scattered within the spectrograph. The geomagnetic 
image motion problem (GIMP) amounted to wavelength-scale shifts of less than 
0.5 pixels (0.9 \AA) for 80\% of the spectra, and was not corrected.

\subsection{HST Broad-band Photometry}
In addition to the 1st order dispersed light, the 0th order undispersed light
was also recorded by the detector array, and provided a broad-band optical/UV
count rate for IP Peg. The passband of the 0th order light has a pivot 
wavelength of 3400~\AA, a full-width at half-peak response of 1900~\AA, and
an effective response of 520 counts s$^{-1}$ mJy$^{-1}$ (accurate to 50\% --
\markcite{Horne93}Horne \& Eracleous 1993; \markcite{Erac94}Eracleous et al.\ 
1994). The 0th order count rates
were converted to an arbitrary magnitude scale via the relation 
\begin{equation}
mag = 12.0 - 2.5\log(count~rate) = 9.28 - 2.5\log\left[{f_{\nu}
(3400{\rm \AA)}\over1\;{\rm mJy}}\right]
\end{equation}
in order to facilitate comparison with ground-based photometry of IP Peg. As 
with the spectra, the photometric measurements have a time resolution of 5.5~s.

\subsection{Ground-based Optical Photometry}
\label{s-phot}
Simultaneous ground-based photometry of IP~Peg was obtained with a TI 
$800\times800$ pixel CCD on the 1-m telescope at the Mount Laguna Observatory.  
Several sub-fields on the CCD were read out into a contiguous image to provide 
a number of comparison stars while minimizing image size and dead time -- for a 
more complete description of the Mount Laguna CCD, see 
\markcite{Shafter93}Shafter, Misselt, \& Veal (1993).  Measurements were made 
through a $V$ filter, with a time resolution of approximately 30~s, between HJD 
= 2448928.77492 and HJD = 2448928.88362.  The images were processed in the 
usual fashion (i.e., bias-subtracted and flat-field-corrected) using standard 
IRAF\footnote{$\,$i.e., the Image Reduction and Analysis Facility operated by 
the 
National Optical Astronomy Observatories} routines.  The reduced stellar count 
rates were converted to magnitudes via aperture photometry.  The magnitude of a 
comparison star was subtracted from that of IP~Peg to remove the effects of 
atmospheric extinction and sky transparency variations.  The ground-based data 
overlaps all of the HST Run~1 observations.
\newpage
\section{Results}
\subsection{The Light Curve of IP Peg during Run 1}
\label{s-lc_xcor}
\begin{figure}[htb]
\epsscale{0.6}
\plotone{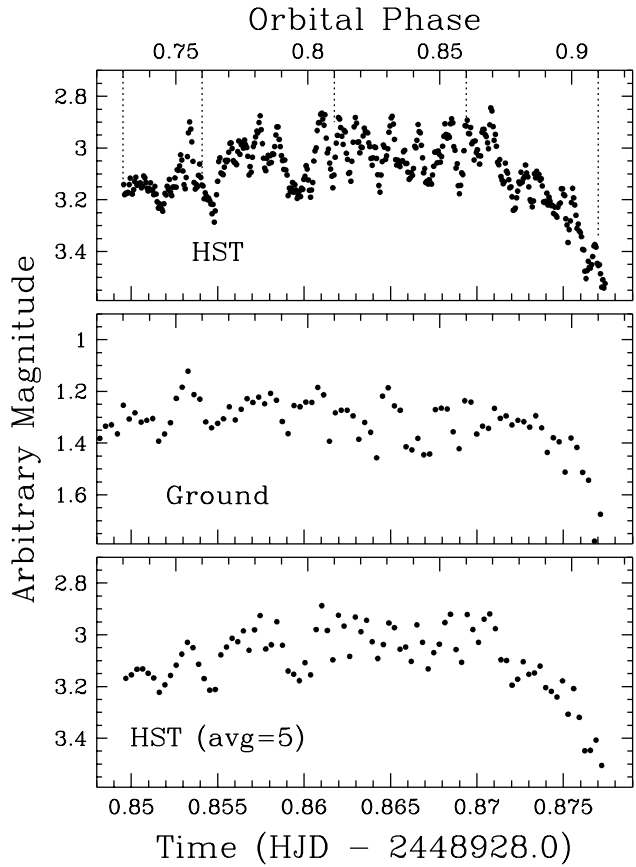}
\caption[]{The ground-based (middle panel) and 0th order HST light curves of 
IP~Peg on 1992~November~2~UT.  The top panel shows the full 5.5 s resolution of 
the HST data while the bottom panel has been averaged over five points to 
simulate the time resolution of the ground-based data.  The horizontal axis at 
the top of each panel is calibrated in orbital phase according to the ephemeris 
of Wolf et al.\ (1993); the horizontal axis at the bottom of each panel is 
calibrated in HJD.  The dotted lines in the top panel mark the limits used to 
construct the phase-resolved spectra described in \S3.2.2 and shown in Figure 4. \label{f-lcs}}
\end{figure}

The HST light curve of IP~Peg obtained from the 0th order passband of the FOS 
observations is shown in the top panel of Figure~\ref{f-lcs}.  
Rapid oscillations with amplitude $\approx 0.3$ mag on timescales of 
$\approx 0.002$ days (about $3$ minutes) are visible.  The middle panel of 
Figure~\ref{f-lcs} shows the simultaneous ground-based light curve of IP~Peg.  
The light curves decline rapidly by $\approx 1$ mag over 0.003~d as the system 
enters eclipse after HJD = 2448928.872.  
For a first,
qualitative comparison of the two data sets, the HST data was binned to
the same time resolution of the ground-based data (bottom panel of 
Figure~\ref{f-lcs}). The two light curves show similarities, the most 
conspicuous
one being the prominent peak near HJD 2448928.853.
Unfortunately, this agreement is achieved at the expense of smearing out the 
short timescale variability seen in the higher time resolution, unbinned HST 
light curve.

In order to gauge the similarity of the two light curves in a quantitative 
manner, {\em Interactive Data Language} (IDL) routines were used to calculate 
the Fourier cross-correlation function of the two data sets.  First, the 
overlapping segments of the light curves from HJD = 2448928.85 until the steep 
ingress to the eclipse at HJD = 2448928.875 were re-binned into equally-spaced 
time series containing 100 points each.  Next, a second order spline function 
was subtracted from each light curve to remove long time scale variations.  
Finally, the cross-correlation was performed and yielded a peak amplitude of 
$\approx +0.54$.  A perfect correlation (e.g., from comparing identical data 
sets) would yield an amplitude of 1.0, so this result indicates a reasonably 
strong positive correlation between the variability in the two light curves.  A 
negative amplitude would have indicated an anti-correlation; that is, one of 
the light curves would show a maximum at the same time the other showed a 
minimum.  In addition, the peak of the cross-correlation function occurs in bin 
0, which indicates that there is no time lag between the variability in the two 
light curves that is larger than the time resolution of the binned data 
($\sim30$ s).  Another measure of the strength of the correlation is obtained 
from the $R$ 
parameter (\markcite{Tonry79}Tonry \& Davis 1979), which is the ratio of the 
amplitude of a given peak to the average amplitude of peaks in the 
cross-correlation function,
\begin{equation}
R = \frac{h}{\sqrt{2}\,\sigma_{a}},
\end{equation}
where $h$~=~the peak height, $\sigma_{a}$~=~the noise level of the 
cross-correlation function, and 
$\sqrt{2}\,\sigma_{a}$~=~the average peak height in the cross-correlation 
function.  The larger the value of $R$, the more significant is the 
correlation; for example, a nearly perfect correlation ($h=0.9$) in the 
presence of noise at the 10\% level ($\sigma_{a}=0.1$) in the cross-correlation 
function would yield $R\approx6.4$, while a correlation peak  
with an amplitude of only twice the noise level would yield $R\approx1.4$.  For 
the IP~Peg light curve cross-correlation, we obtain a correlation peak with 
$R\approx3.6$, corresponding to a peak with amplitude of about 5 times the 
noise level in the cross-correlation function.

Although the behavior of the HST 0th order and ground-based $V$ light curves of 
IP~Peg is fairly strongly correlated -- which indicates that the source of the 
short timescale variability radiates over a wavelength range of 
$\approx3400$--$5500$ \AA\ -- it should be kept in mind that, as mentioned 
above, the shortest timescale oscillations cannot be detected in the lower time 
resolution ground-based data.  Thus, there may yet be wavelength dependent 
differences between the 0th order and $V$ filter bandpasses at the very short 
timescales of the flickering seen in the unbinned HST light curve.

We also calculated the cross-correlation of the 0th order undispersed light 
curve and a spectrophotometric ``flux curve'' of the UV continuum 
($\lambda\approx2000$--$2100$ \AA) from the 1st order dispersed light (see 
\S\ref{s-fc}).  However, no significant correlation was found (the highest peak 
in the cross-correlation function has amplitude $\lesssim0.15$ and is 
comparable to the typical amplitude of random peaks).  This implies that the 
source of the continuum radiation in IP Peg is different in the UV than in the 
optical.  

\subsection{The Ultraviolet Spectrum of IP Peg during Run 1}
\begin{figure}[hbtp]
\epsscale{0.85}
\plotone{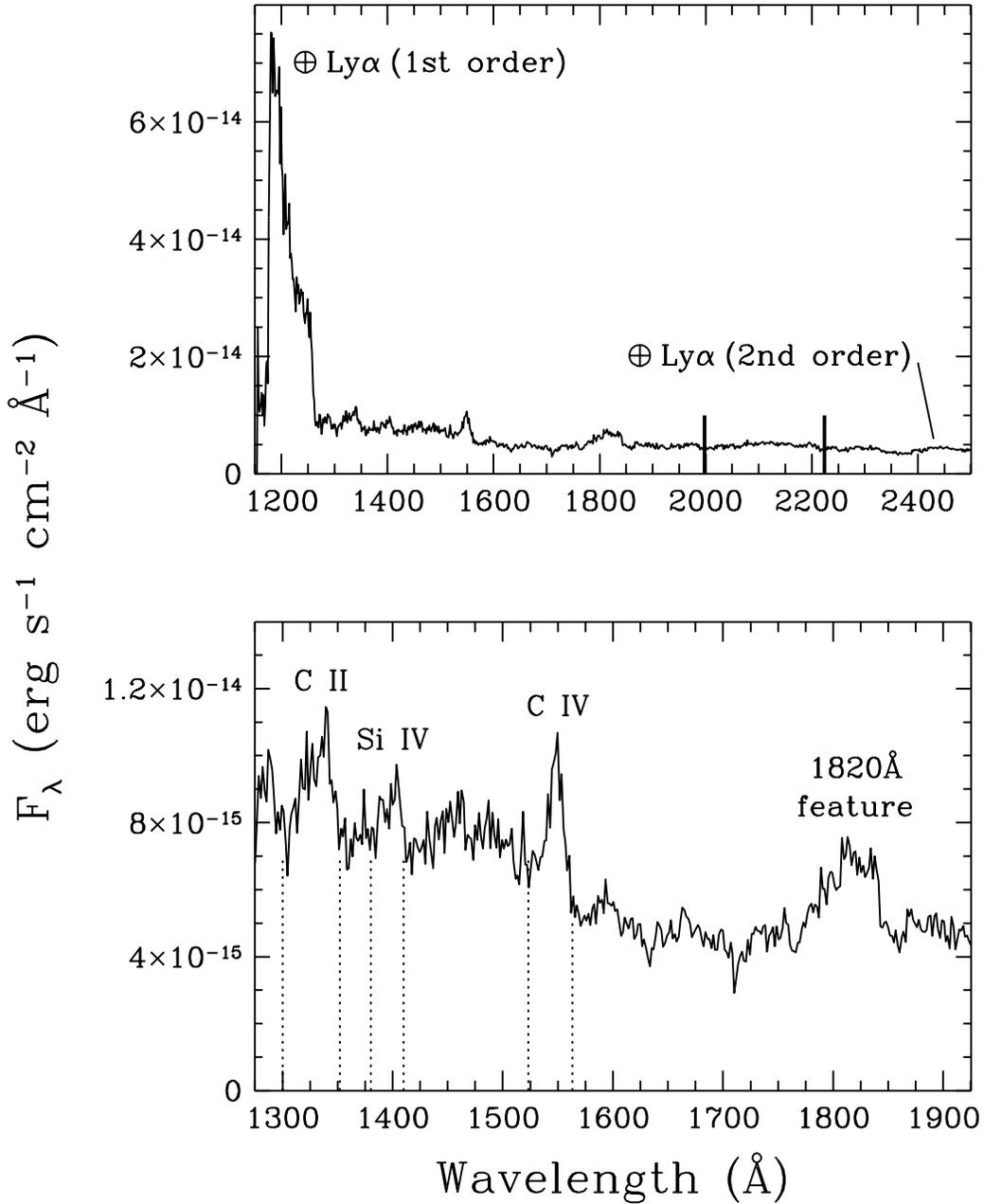}
\caption[]{The average HST/FOS spectrum of IP Peg from Run 1 on 1992 November 2 
UT.  The top panel shows the entire spectrum with the positions of the first- 
and second-order (geocoronal) Ly$\alpha$ emission indicated.  The dark vertical 
bars near 2000 \AA\ and 2200 \AA\ mark the limits of the region from which the 
continuum flux curves used in the cross-correlation analysis were obtained.  
The bottom panel is an enlargement of the central region of the spectrum 
showing the emission lines of carbon (C~{\sc ii} $\lambda1335$, C~{\sc iv} 
$\lambda1550$) and silicon (Si~{\sc iv} $\lambda1400$), as well as the unusual 
feature at $\lambda\approx 1820$ \AA.  The dotted lines mark the limits of the 
wavelength regions from which the emission line flux curves used in the 
cross-correlation analysis were obtained.  \label{f-spec}}
\end{figure}

The average HST spectrum of IP~Peg is shown in Figure~\ref{f-spec}.  
The spectrum displays a number of emission lines of various elements, primarily 
carbon (\ion{C}{2} $\lambda1335$, \ion{C}{4} $\lambda1550$) and silicon 
(\ion{Si}{4} $\lambda1400$).  There is also an unusual and unidentified 
emission-like feature centered around $\lambda\approx1820$ \AA. 

\subsubsection{The 1820 \AA\ Feature}
This feature in the ultraviolet spectrum is plateau-like.  It peaks at 
$\approx1820$ \AA, showing a width of $\approx75$ \AA\ and an intensity of 
about $2\times10^{-15}$ ergs s$^{-1}$ cm$^{-2}$ \AA$^{-1}$ above the local 
continuum level, comparable to that of the carbon and silicon lines.  Its long 
wavelength edge is very sharp.  The IUE spectrum of IP~Peg obtained by 
\markcite{Szkody87}Szkody (1987) over similar wavelength range, orbital phase, 
and total exposure time does {\em not} display this feature.  
It is weakly present in spectra obtained with an identical FOS configuration 
during the eclipse of IP~Peg in Run~3 of our 1992~November HST observations and 
is very strong in follow-up HST observations of IP~Peg obtained during decline 
from an outburst in 1993~May; however, 
it is completely gone in HST/FOS spectra taken during an outburst
(\markcite{Bap94}Baptista et al. 1994).  
We have examined the possible origin of this feature as (a) a data artifact, 
(b) a superposition of emission lines, or (c) a gap between broad absorption 
bands.

\begin{figure}[tbp]
\plotfiddle{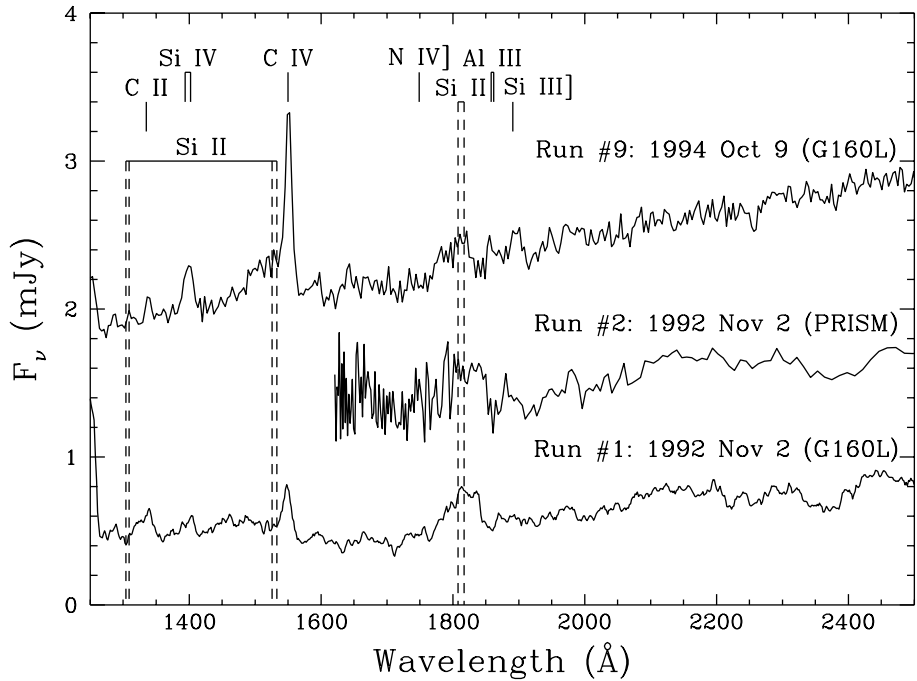}{3.25in}{-90}{50}{50}{-195}{280}
\caption[]{HST/FOS spectra of IP Peg obtained on the same day using different 
optical elements (Runs 1 and 2), and on a different day two years later 
(following the decline from an outburst) using the same optical elements as in 
Run 1 (Run 9).  The flux scale refers to the bottom spectrum (Run 1) and the 
remaining spectra have been separated by 0.75 mJy.  The positions of several 
spectral lines are indicated. \label{f-1820}}
\end{figure}

(a) Incorrect flat-fielding of FOS images can produce erroneous features in the 
spectra with characteristics similar to those of the 1820 \AA\ feature 
(\markcite{Leith95}Leitherer 1995).  In addition, the G160L disperser is 
subject to large ($> 5$\%) inaccuracies in the time-dependent behavior of its 
flat-field.  An inspection of the flat-field in that spectral region, however, 
did not reveal any irregularities.  Figure~\ref{f-1820} shows the average 
spectrum of IP Peg from Runs 1, 2, and 9.
The last two of these were HST/FOS 
observing runs similar in nature to Run 1 as it is described in 
\S\ref{s-hst_obs}.  The spectra in Run 2 (Baptista et al. 1994) were obtained 
on the same day as Run 1, but using the PRISM rather than the G160L grating.  
Run 9 (which will be described in detail in a future paper) was obtained with 
the same grating as Run 1, but $\approx2$ yr later following the decline from 
an outburst of IP Peg.  The 1820 \AA\ feature is visible in all three of these 
spectra.  The fact that it appears in both the PRISM and G160L spectra rules 
out an instumental artifact because in these two configurations the spectra 
fall on different locations on the FOS/BLUE photocathode and are recorded by 
different parts of the diode array.

(b) Although \ion{Si}{2} has lines at 1808 \AA\ and 1817 \AA, the width and 
oddly-shaped profile of the 1820 \AA\ feature is not expected from simple line 
emission.  If the \ion{Si}{2} doublet were contributing a large amount of the 
flux, then we might expect the profile of the 1820 \AA\ feature to look more 
like the strong (unresolved) doublet of \ion{Si}{4} near 1400 \AA.  In 
addition, \ion{Si}{2} has two other doublet-producing transitions to the ground 
state (at 1304, 1309 \AA\ and 1526, 1533 \AA) that we would expect to see if 
the 1808, 1817 \AA\ doublet was strong.  These additional doublets are not 
visible in the spectra of IP Peg (see Fig.~\ref{f-1820}).

(c) A feature similar to the one at $\approx1820$ \AA\ in the spectrum of 
IP Peg is observed in the spectrum of the cataclysmic variable OY~Car.  In the 
case of OY~Car, the feature is thought to be a region of continuum that appears 
to be in emission because it is sandwiched between two adjacent absorption 
bands of the ``\ion{Fe}{2} curtain'' discussed by \markcite{Horne94}Horne et 
al.\ (1994).  Yet, there are a number of differences between the spectra of 
IP~Peg and OY~Car.   For example, while there is some indication of the 
presence of an absorption band on the short wavelength side of the 1820 \AA\ 
feature in IP~Peg, the long wavelength side is relatively smooth and constant 
as far out as the beginning of the 2nd order geocoronal Ly$\alpha$ emission.  
The latter region, at least, is quite continuum-like, with no apparent 
absorption bands.  There does seem to be a rather abrupt drop in the (apparent) 
continuum level from the short-wavelength side of the \ion{C}{4} $\lambda1550$ 
line to the long-wavelength side, and the flux level of the 1820 \AA\ feature 
is very close to that of the continuum level shortward of \ion{C}{4}.  This 
suggests the possibility that the flux level across the entire spectral region 
from $\approx1550$ \AA\ to $\approx1750$ \AA\ may be suppressed by absorption 
bands of the \ion{Fe}{2} curtain.  The fact that the spectrum of IP Peg does 
not resemble the spectrum of OY Car very closely does not necessarily preclude 
the presence of an Fe curtain of some sort.  It is possible that the ionization 
structure of the curtain is different in the two stars (e.g., due to a 
difference in the luminosity of the background source or a change in the column 
density caused by a difference in inclination angle).  The absence of the 1820 
\AA\ feature in the IUE spectrum presented by Szkody (1987) may also be related 
to changes in the ionization structure of the absorbing curtain.

Although the actual origin of the 1820 \AA\ feature in the spectrum of IP~Peg 
is not yet certain, we believe that data artifacts and contribution from 
underlying emission lines can be ruled out.  Despite the cosmetic differences 
between the spectrum of IP Peg and OY Car, this leaves a manifestation of the 
\ion{Fe}{2} curtain as the most likely cause for the 1820 \AA\ feature.

\subsubsection{The Emission Lines}
The fluxes and equivalent widths of the prominent emission lines in the average 
HST spectrum of IP~Peg were measured by direct integration of the pixel 
intensities between two manually-selected endpoints.  The results are shown in 
Table~\ref{tab1}.  In each case, the line parameters were measured from the 
spectrum three separate times; the tabulated results are the average values and 
standard deviation of these three trials.  The uncertainties in the averages of 
the measured line centers are all much smaller than 1 \AA.  
Also listed in Table~\ref{tab1} are the flux and equivalent width of the 
\ion{C}{4} $\lambda1550$ line from the IUE spectrum of IP~Peg obtained over a 
\begin{table}
\dummytable\label{tab1}
\end{table}
\begin{figure}[tb]
\plotfiddle{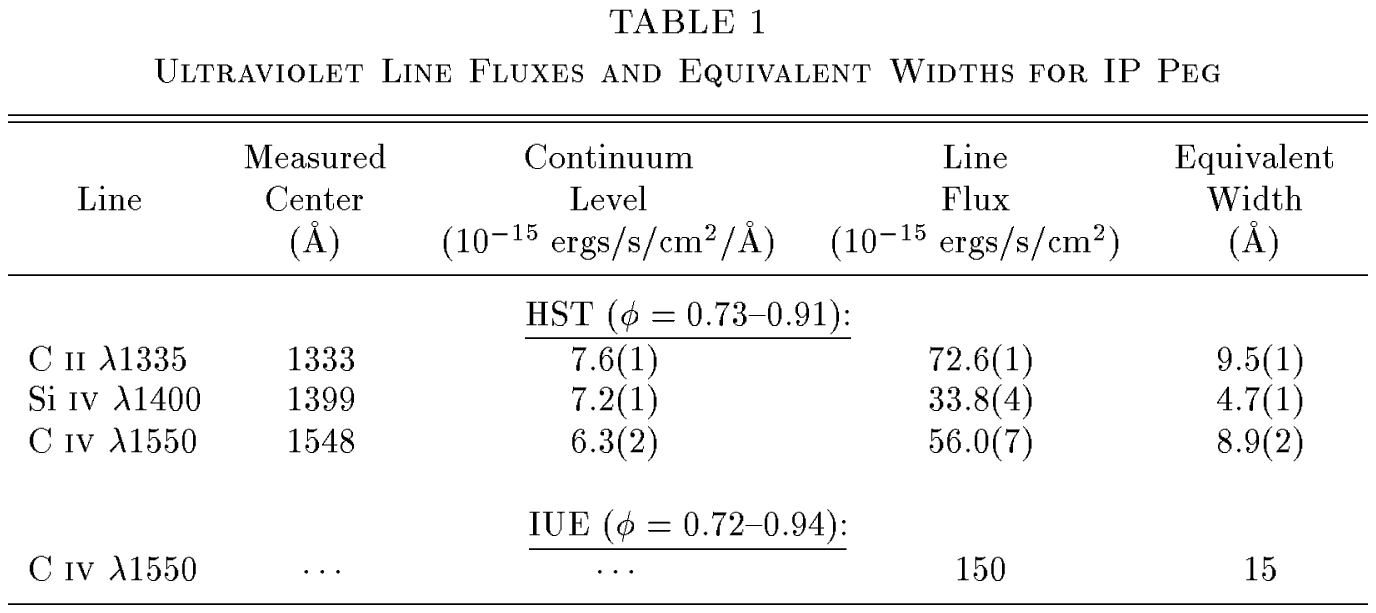}{2.5in}{0}{100}{100}{-305}{-300}
\end{figure}
similar range of orbital phase as the HST spectrum, but six years earlier, on 
1986 August 22 UT (= JD 2446664; Szkody 1987).  The flux of the \ion{C}{4} line 
in the IUE spectrum is almost three times that in the HST spectrum, and the 
equivalent width is about twice as large in the former.  IP Peg was in a 
quiescent state during both of these observations: the {\em Variable Star 
Network} (http://www.kusastro.kyoto-u.ac.jp/vsnet/) long-term light curve for 
IP Peg shows it to have been between outbursts, with $V\approx15$--$16$.  Also, 
the continuum levels in the IUE and HST spectra are comparable, at 
$F_{\lambda}\approx0.5$--$1.0\times10^{-14}$ erg s$^{-1}$ cm$^{-2}$ \AA$^{-1}$.
Because of the large time difference between the two spectra, however, it is 
not clear whether these differences reflect the range of orbit-to-orbit 
variability of the emission lines or an actual long term change in the emission 
characteristics of IP~Peg. 

\begin{figure}[tb]
\plotfiddle{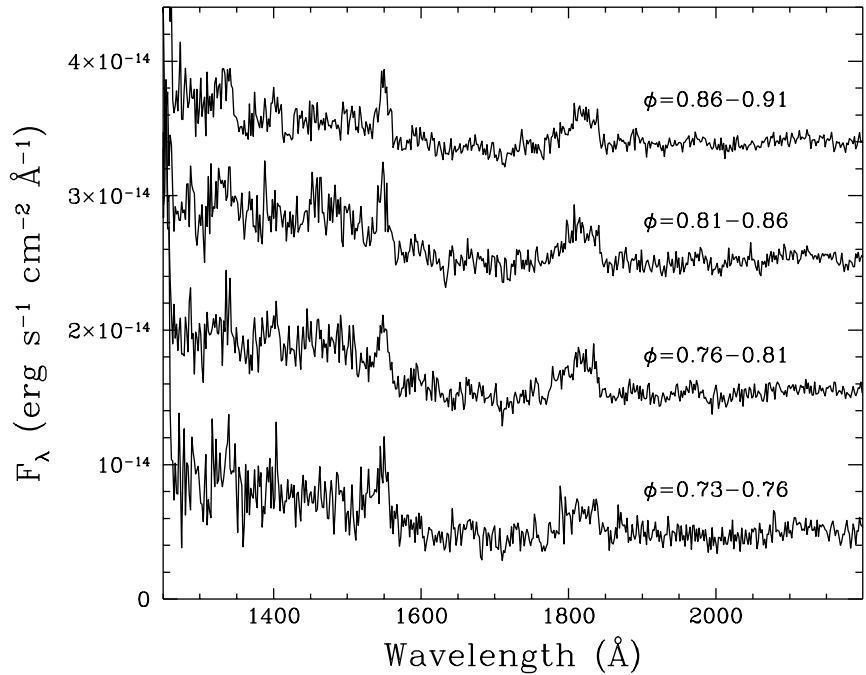}{3.35in}{-90}{50}{50}{-225}{280}
\caption[]{The phase-resolved spectra of IP Peg.  The flux scale refers to the 
bottom spectrum and $1.0\times10^{-14}$ erg s$^{-1}$ cm$^{-2}$ \AA$^{-1}$ has 
been cumulatively added to each spectrum above it. \label{f-phase_spec}}
\end{figure}

Since changes in the nature of its spectrum should be expected as IP~Peg goes 
into eclipse, the averaged spectrum was divided into several successive 
sub-spectra: one of width 0.03 (starting at $\phi=0.73$) and three covering 
phase bins of width 0.05 (ending at $\phi=0.91$).  These phase-resolved spectra 
of IP Peg are shown in Figure~\ref{f-phase_spec}.  The line fluxes and 
equivalent widths in the earliest (out of eclipse) and the latest (beginning of 
eclipse) of the spectra with phase bin widths of $\Delta\phi = 0.05$ were 
measured as described above; the results are shown in Table~\ref{tab2}.  With 
the exception of the \ion{Si}{4} line, which shows an essentially unchanged 
equivalent width from one spectrum to the other, the widths of the carbon lines 
are lower in the out of eclipse spectrum by a factor of $\approx2$--$4$.  In 
both of the carbon lines, the flux level at the beginning of eclipse is larger 
than that at out of eclipse phases.  The \ion{Si}{4} line, on the other hand, 
has a larger flux outside of eclipse.  This suggests that the \ion{Si}{4} line 
may be originating in a narrower, more central region in the disk than the 
carbon lines; hence, the former is being eclipsed more than the latter.
\begin{table}
\dummytable\label{tab2}
\end{table}
\begin{figure}[tb]
\plotfiddle{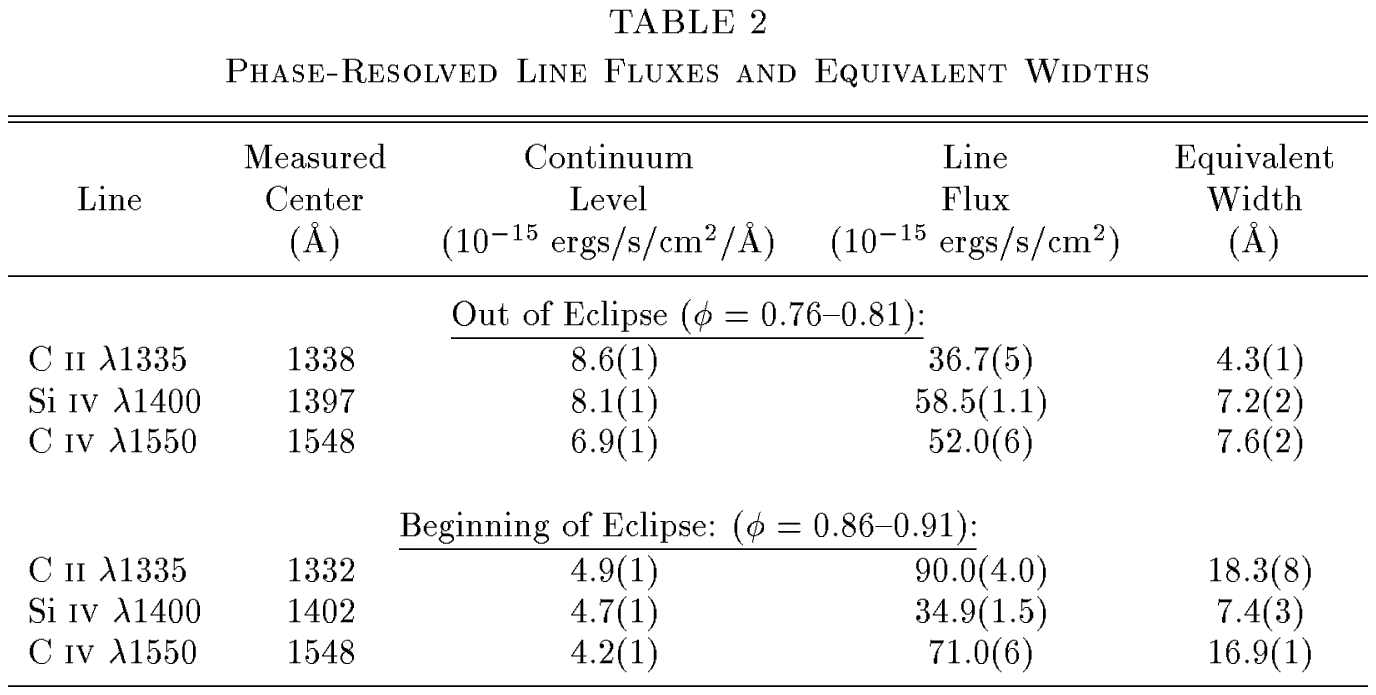}{2.6in}{0}{100}{100}{-305}{-300}
\end{figure}

\subsubsection{The Flickering Spectrum}
\label{s-flick_spec}
\begin{figure}[htb]
\epsscale{0.55}
\plotone{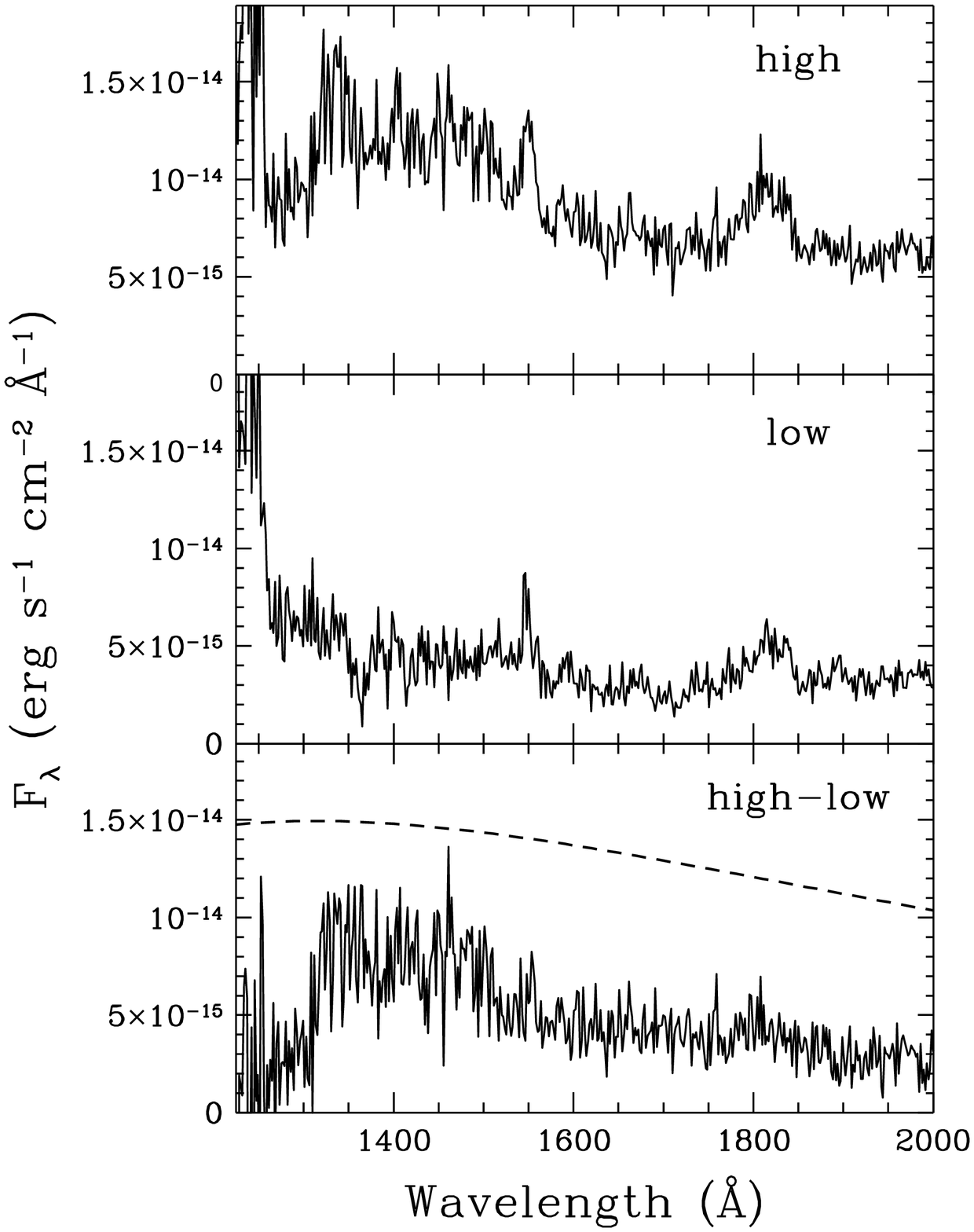}
\caption[]{The top panel shows the average ``high'' spectrum of IP Peg 
obtained at the peaks of the flickering.  The middle panel shows the average 
``low'' spectrum obtained between flickers.  The bottom panel is the difference 
of the high and low spectra, showing the wavelength-dependence of the 
flickering amplitude.  The dashed line in the bottom panel is an 
arbitraily-scaled blackbody curve with $T=22,000$ K for comparison. 
\label{f-flicker_spec}}
\end{figure}

We determined the mean flux level of each HST spectrum in the wavelength 
interval 1300--2200 \AA, then calculated the average and standard deviation 
($\sigma_{mean}$) of the mean flux levels.  All of the spectra with mean flux 
levels greater than $1\sigma_{mean}$ above the average were combined to give 
the spectrum of IP Peg at the peaks of the UV flickering (henceforth, the high 
spectrum), while those with mean levels less than $1\sigma_{mean}$ below the 
average were combined to give the spectrum between flickers (henceforth, the 
low spectrum).  The low spectrum was then subtracted from the high spectrum to 
obtain the wavelength-dependence of the UV flickering amplitude (henceforth, 
the flickering spectrum).  These three spectra are shown in 
Figure~\ref{f-flicker_spec}.

The \ion{C}{4} line is present with comparable strength in both the high and 
low spectra.  The \ion{C}{2} and \ion{Si}{4} lines, on the other hand, are only 
apparent in the high spectrum.  This suggests a possible relationship between 
the \ion{C}{2} and \ion{Si}{4} emission sources that is not shared by the 
source of the \ion{C}{4} emission.

The 1820 \AA\ feature is visible in both the high and low spectra, but is 
almost entirely gone in the flickering spectrum.  The high and low spectra were 
constructed from approximately equal-number samples of more-or-less randomly 
distributed spectra in the original continuous series of spectra.  Thus, a 
feature which is equally visible in all of the original spectra (e.g., due to 
the effect of an \ion{Fe}{2} curtain) should be present with equal strength in 
both the high and low spectra.  This would cause it to be totally subtracted in 
the flickering spectrum, as is observed for the 1820 \AA\ feature.

There is a region of excess flux in the flickering spectrum between 
$\approx1300$ \AA\ and $\approx1500$ \AA, which indicates that a large fraction 
of the UV flickering originates in this wavelength region.  An obvious 
conclusion is that this excess flux represents the contribution of the bright 
spot.  A blackbody with $T\approx22,000$ K would peak near the center of this 
feature, at $\lambda\approx1400$ \AA.  This temperature is comparable to that 
determined for the bright spot in IP Peg from IUE observations (Szkody 1987); 
however, attempts to fit a blackbody curve to the HST flickering spectrum were 
unsuccessful because of the narrowness of the excess flux feature and the sharp 
drop in flux shortward of $\approx1300$ \AA.  The dashed line in the bottom 
panel of Figure~\ref{f-flicker_spec} shows an arbitrarily-scaled blackbody 
curve for $T=22,000$ K.  Although the slope of the curve longward of 
$\approx1500$ \AA\ is a reasonable match to the data, the shape of the curve 
deviates significantly from the data at shorter wavelengths.  The flickering 
spectrum does not appear to be adequately modelled by a simple blackbody, 
suggesting that a more sophisticated model is appropriate.  For example, a disk 
atmosphere spectrum might account for the depression shortward of 1300 \AA\ 
as a broad Ly$\alpha$ absorption line.  For now, however, the origin of the 
excess flux and short-wavelength flux drop-off in the wavelength-dependence of 
the UV flickering amplitude are unclear.  

\subsubsection{Comparison of Line and Continuum Flickering}
\label{s-fc}
Spectrophotometric ``flux curves'' were constructed from the IP~Peg spectra by 
averaging (with equal weights) the pixel values in a specific wavelength range 
(spectral bin) in each spectrum.  Because the wavelength scale is linear to a 
high degree of accuracy, this average with equal weights is valid even for a 
reasonably wide spectral bin and for bins taken from different regions of the 
spectra.  The pixel and wavelength ranges used, along with the mean values of 
the resultant flux curves, are listed in Table~\ref{tab3}.  The primary 
wavelength regions were centered around the \ion{C}{2}, \ion{Si}{4}, and 
\ion{C}{4} emission lines, as well as around two regions of pure continuum.  We 
also created flux curves from small continuum regions located adjacent to both 
endpoints of the emission line spectral bins.  
\begin{table}
\dummytable\label{tab3}
\end{table}
\begin{figure}[tb]
\plotfiddle{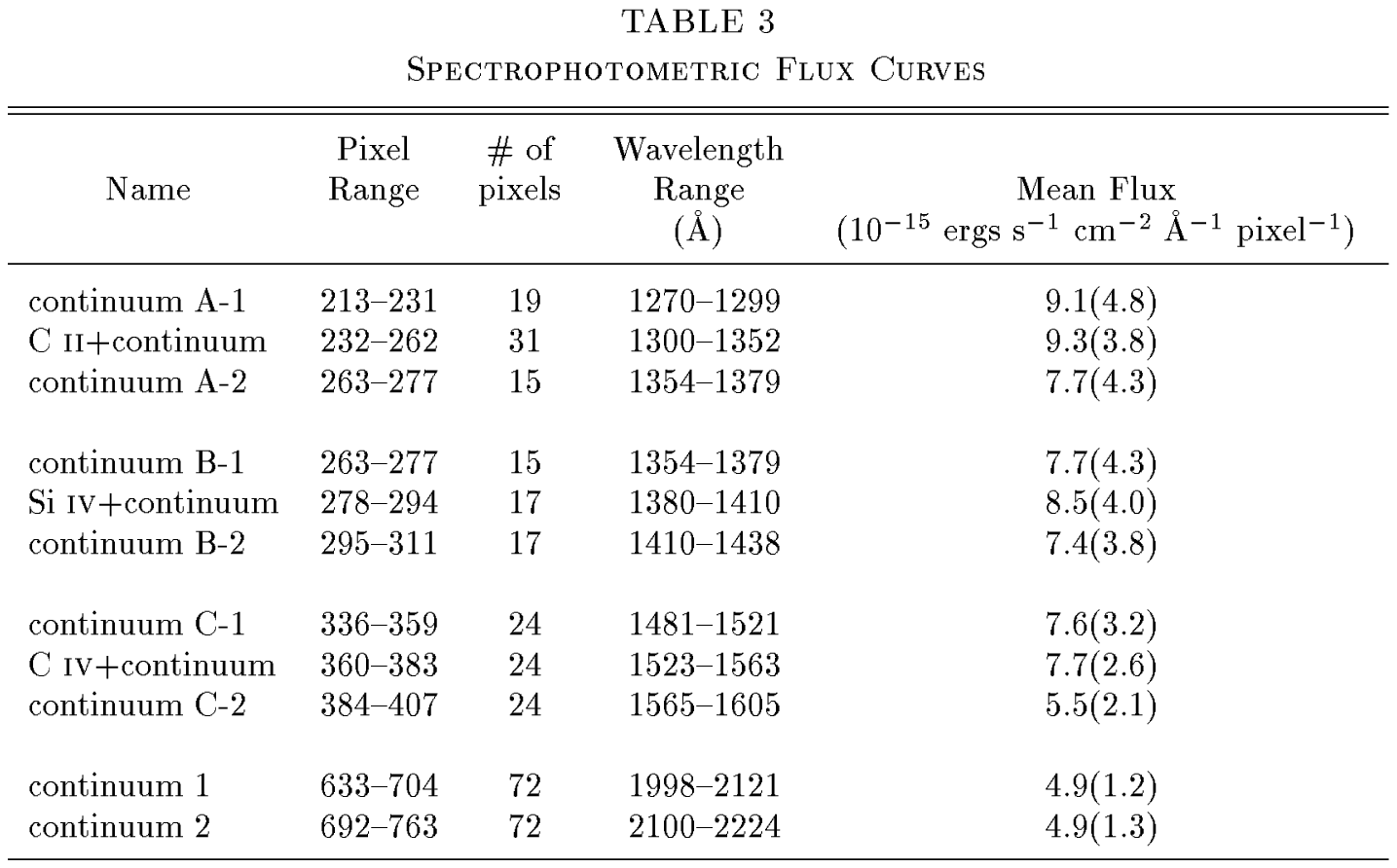}{3.5in}{0}{100}{100}{-305}{-260}
\end{figure}

The wavelength ranges of the small continuum regions adjacent to the lines were 
chosen to minimize contamination from other spectral lines.  In general, these 
continuum regions are line-free, with the exception that \ion{O}{5} 
$\lambda1371$ falls in A-2/B-1 and \ion{N}{4}] $\lambda1487$ falls in C-1.  
However, we inspected the average spectrum of IP Peg and found no evidence for 
the presence of these lines above the scatter in the continuum level.  In order 
to isolate the flickering behavior of the emission lines from that of the 
underlying continuum, the pair of continuum flux curves for the regions 
adjacent to each line were averaged together and subtracted from the flux curves of the corresponding line regions.  The mean values and $1\sigma$ standard 
deviations of the averaged continuum flux curves and the continuum-subtracted 
line flux curves are shown in Table~\ref{tab4}.
\begin{table}
\dummytable\label{tab4}
\end{table}
\begin{figure}[tb]
\plotfiddle{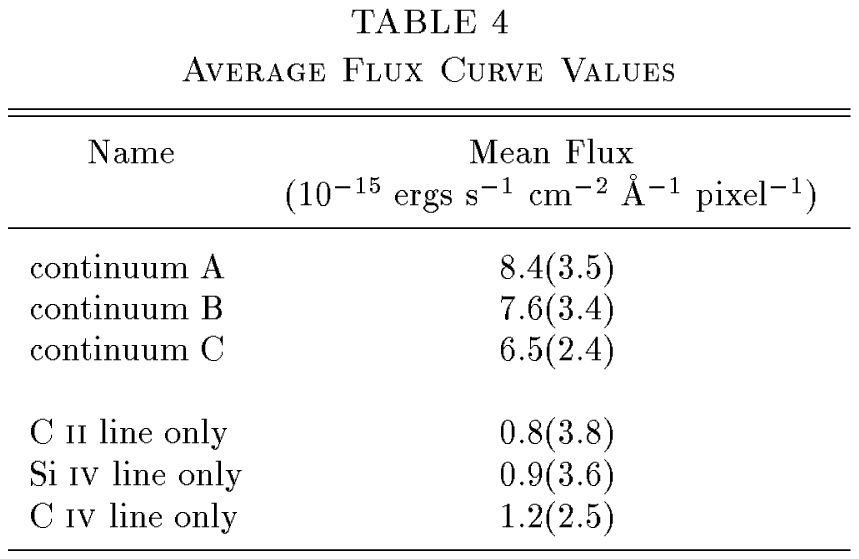}{2.0in}{0}{100}{100}{-305}{-315}
\end{figure}

The IP~Peg flux curves were cross-correlated as in \S\ref{s-lc_xcor} according 
to the pairings listed in Table~\ref{tab5}.  The first few cases test the 
continuum region flux curves for contamination from non-continuum flickering, 
in order to ensure that the regions used are truly representative of the 
continuum behavior.
\begin{table}
\dummytable\label{tab5}
\end{table}
\begin{figure}[tbp]
\plotfiddle{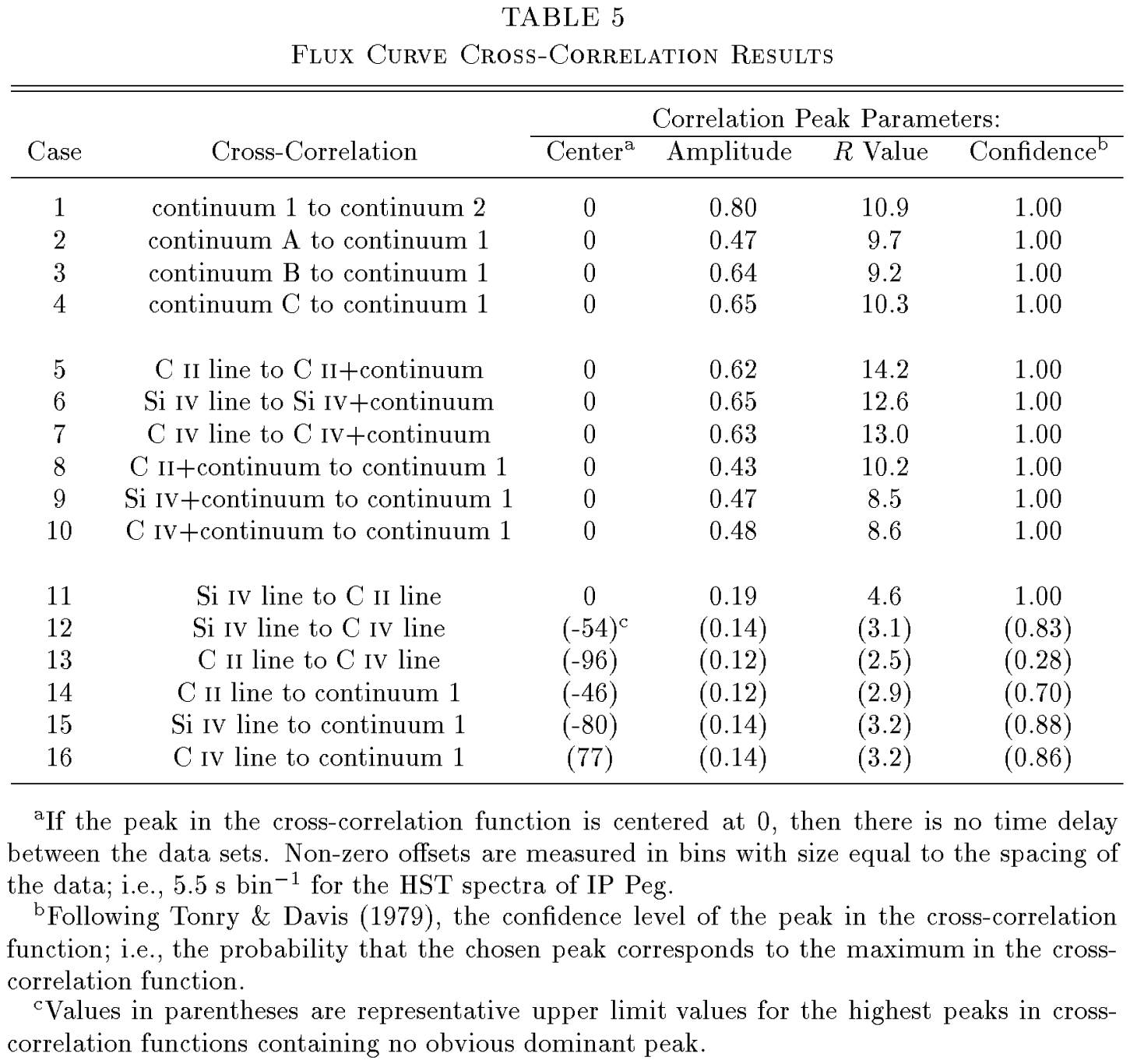}{6.0in}{0}{100}{100}{-305}{-110}
\end{figure}

\begin{description}
\vspace{-9pt}

\item[Case 1:] These two continuum sections were chosen from a line-free region 
of the spectrum.  Their high degree of correlation implies that there is little 
or no contamination from flux variability other than random noise differences 
between the two continuum regions; that is, these regions appear to be pure 
continuum.

\item[Cases 2, 3, 4:] These cases compare one of the pure continuum samples 
with the flux curves of the continuum regions adjacent to the lines, which were 
subtracted from the line flux curves to remove the continuum contribution.  
The average continuum flux curves used to background-subtract the emission lines are highly correlated with the pure continuum.

\vspace{-9pt}
\end{description}

The next six cases provide additional tests of the cross-correlation procedure 
by comparing flux curves whose cross-correlation has a predictable outcome.

\begin{description}
\vspace{-9pt}

\item[Cases 5, 6, 7:] These cases are highly correlated, as expected due to the 
presence of the line flickering component in both the line-only and 
line-plus-continuum flux curves.

\item[Cases 8, 9, 10:] These cases are also highly correlated, as expected due 
to the presence of the continuum flickering component in both the 
line-plus-continuum and continuum-only flux curves.

\vspace{-9pt}
\end{description}

Finally, the remaining cases deal with the actual comparisons of interest here, 
from line-to-line and line-to-continuum.

\begin{description}
\vspace{-9pt}

\item[Cases 11, 12, 13:] These cases are expected to show a high degree of 
correlation if the different emission lines are varying in the same fashion.  
This would imply that the lines originate from the same region in the system.  
The fact that the \ion{C}{2} and \ion{Si}{4} lines {\em are} correlated with 
each other (Case 11), while neither is correlated with the \ion{C}{4} line 
(Cases 12 and 13), suggests two possible conclusions: (a) the \ion{C}{2} and 
\ion{Si}{4} lines actually are correlated with the \ion{C}{4} line, but at a 
level that is small compared to the noise in the flux curves, or (b) the 
\ion{C}{2} and \ion{Si}{4} lines originate in the same physical region of the 
system and it is a different region than that in which the \ion{C}{4} line 
originates.  We favor the second possibility, which is also supported by the 
relative strengths of the emission lines in the high and low flickering spectra 
of IP Peg (see \S\ref{s-flick_spec}), as well as the fact that the relatively 
high ionization energy of \ion{C}{4} ($\chi=64.5$ eV) suggests that this line 
is likely to originate in a more energetic region than either \ion{Si}{4} 
($\chi=45.1$ eV) or \ion{C}{2} ($\chi=24.4$ eV).

\item[Cases 14, 15, 16:] These cases are expected to show a high degree of 
correlation only if the lines and continuum are varying together.  Since none 
of the cases shows significant correlation, it must be true that the lines and 
continuum behave independently of each other.

\vspace{-10pt}
\end{description}

\begin{figure}[htb]
\plotfiddle{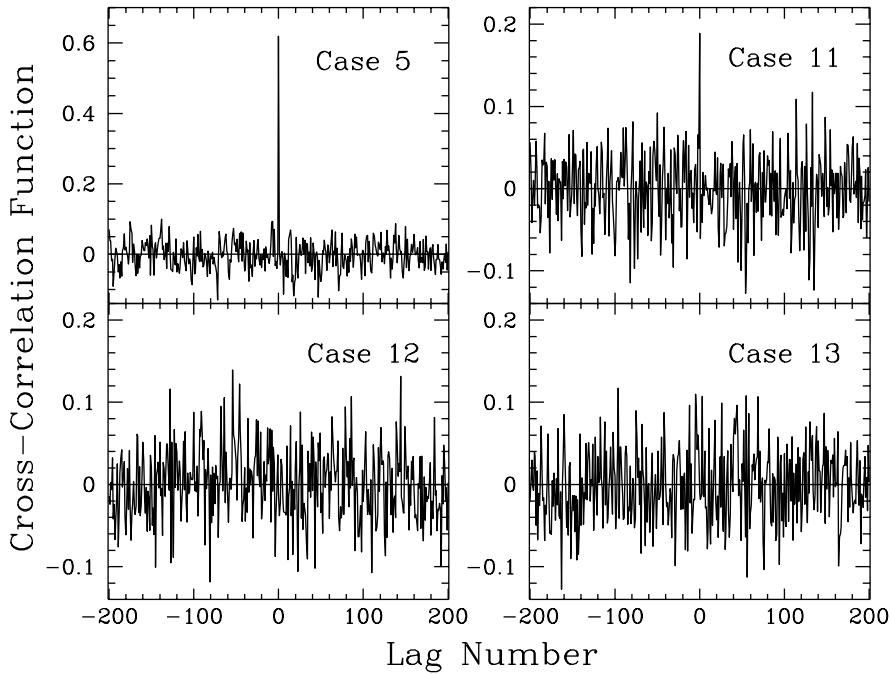}{3.5in}{-90}{50}{50}{-200}{285}
\caption[]{The cross-correlation functions for Cases 5, 11, 12, and 13 (see 
Table 5).  Only Cases 5 and 11 exhibit a significant correlation. 
\label{f-xcor1}}
\end{figure}

The amplitudes and $R$ values of the peak in Case 11 are not much larger than 
those of the peaks in the remaining higher-numbered cases.  Hence, one might 
ask why the former peak is considered significant, while the latter are not?  
The rightmost column in Table~\ref{tab5} lists a confidence level (which was 
calculated as in Tonry \& Davis 1979) corresponding to the selected peak in the 
cross-correlation function.  Through Case 11, this confidence level is 1.00, 
indicating that the chosen peak is the true maximum in the cross-correlation 
function.  While it is not difficult to accept that the peak in Case 13, with a 
confidence level of only 0.28, is spurious, the remaining cases beyond Case 11 
have confidence levels of $\approx0.7$--$0.9$.  We might normally be inclined 
to accept a result of which we are 80\% confident; however, a ``normal'' 
interpretation of the confidence level is not applicable in this case.  The 
error function for the cross-correlation peak changes rapidly for small values 
of $R$ then levels out at larger values of $R$ (e.g., see Fig.~14 in Tonry \& 
Davis 1979).  Consequently, while a confidence level of 1.00 does indicate that 
the correlation is highly significant, a spurious peak in the cross-correlation 
function can produce a confidence level only a few percent smaller.  This is 
best illustrated by comparing plots of the cross-correlation functions for 
several cases that have different peak amplitudes, $R$ values, and confidence 
levels.  Figure~\ref{f-xcor1} shows the cross-correlation functions of Cases 5, 
11, 12, and 13.  The strong peak in Case 5 has a high amplitude, large $R$ 
value, and confidence level of 1.00 (see Table~\ref{tab5}).  The height of the 
peak in Case 11 (which also has a confidence level of 1.00) is still well in 
excess of the noise fluctuations in the rest of the curve, despite its 
relatively smaller amplitude and $R$ value.  The remaining two plots, on the 
other hand, do not contain an obvious dominant peak.  The cross-correlation 
function of Case 12, whose highest peak has a confidence level of $\approx0.8$, 
does not contain any single peak that is preferrable over others and is 
visually indistinguishable from Case 13, whose highest peak has only a low 
confidence level ($\approx0.3$).

\section{Discussion and Conclusions}
The flickering in the simultaneous ground-based optical ($V$-band) and HST 
broad-band (undispersed 0th order) light curves of IP~Peg shows a fairly strong 
degree of correlation.  This suggests that the source of the optical continuum 
radiation emits over a wavelength interval of at least $\approx3400$--$5500$ 
\AA; however, the difference in time resolution between the two data sets 
(i.e., the low time resolution of the ground-based data) prevents a more 
detailed comparison of the flickering at these wavelengths.  On the other hand, 
the optical continuum is not correlated with the UV continuum, which implies 
that the continuum radiation has a different source shortward of 
$\lambda\sim2200$ \AA.  None of the UV emission line flickering shows a 
significant degree of correlation with that of the UV continuum, which 
indicates that the lines and continuum are produced by separate mechanisms 
and/or in different locations in the IP~Peg system.  

In both of the emission lines of carbon (\ion{C}{2} and \ion{C}{4}), the line 
fluxes at the beginning of eclipse ($\phi$ = 0.86--0.91) are larger than those 
out of eclipse ($\phi$ = 0.76--0.81).  The \ion{Si}{4} line, on the other hand, 
has a larger flux outside of eclipse.  Also, the equivalent widths of the 
carbon lines increase by factors of $\approx$2--4 between the out of eclipse 
and beginning of eclipse spectra, but the equivalent width of the silicon line 
remains approximately constant.  This suggests that the \ion{Si}{4} emission 
may be originating in a narrower, more central region in the disk than the 
carbon lines; hence, the former is being eclipsed more than the latter.  An 
equivalent statement is that there may be carbon, but not silicon, emission 
that is visible as the system goes into eclipse, possibly originating in a 
vertically-extended corona/wind from the disk.

While the \ion{C}{2} and \ion{Si}{4} emission line flickering is moderately 
correlated, neither is correlated with the flickering of the \ion{C}{4} line.  
This suggests that the \ion{C}{4} line originates from a different physical 
location in the IP~Peg system than the other two lines.  This conclusion is 
also supported by a comparison of the relative strengths of the emission lines 
in the high and low flickering spectra of IP Peg.  Yet, it appears to 
contradict somewhat the flux level and equivalent width behavior of the 
emission lines going into eclipse, which shows that the two carbon lines behave 
similarly to each other and differently from the silicon line.  This dilemma 
could be resolved if, for example, the vertical structure above and below the 
accretion disk is stratified into a hot corona with an equilibrium 
configuration close to the disk surface, and an outflowing wind that dominates 
further away from the disk.  A similar model was discussed by 
\markcite{Meyer94}Meyer \& Meyer-Hofmeister (1994) as it relates to the 
dissipation of the inner disk in dwarf novae via a coronal siphon flow.  If the 
\ion{Si}{4} emission originates in the corona near the disk surface, the 
\ion{C}{2} originates in the upper corona, and the \ion{C}{4} emission 
originates in the wind, then both carbon emitting regions could be visible 
going into eclipse while the silicon emission region is obscured.  At the same 
time, the \ion{C}{2} line is still forming in the same overall region (the 
corona) as the \ion{Si}{4} line, so these two lines might vary in the same 
fashion.  The \ion{C}{4} line, on the other hand, forms in a different region 
(the wind), and varies differently.

The results presented here are based on observations obtained when the bright 
spot in IP Peg was most visible.  A number of recent simulations of accretion 
disk structure have shown that the region where the accretion stream impacts 
the disk is likely to be overly dense and vertically extended 
(\markcite{Livio86}Livio, Soker, \& Dgani 1986, \markcite{Hirose91}Hirose, 
Osaki, \& Mineshige 1991, \markcite{Arm96}Armitage \& Livio 1996).  Thus, it 
must be kept in mind that the flickering behavior of the lines and/or continuum 
is likely to be influenced by the presence of the bright spot.  For example, 
the lack of correlation between the flickering in the UV and optical continua 
may be the result of the bright spot contributing more flux in one or the other 
of these wavelength regimes.  However, the discrepancy in estimated 
temperatures for the bright spot in IP Peg (see \S\ref{s-intro}) makes it 
difficult to determine which wavelength regime is more likely to be affected by 
the spot.  Also, the fact that the IP~Peg emission lines in the UV and optical 
are brighter at phases away from the pre-eclipse hump (Marsh 1988, Baptista et 
al. 1994) suggests that the bright spot may partially obscure emission line 
sources in the inner disk.  Doppler tomography performed on the UV emission 
lines in IP~Peg would be most helpful in revealing the relative locations of 
the UV emission regions.  

\acknowledgments
The observations were made with the NASA/ESA {\em Hubble Space Telescope}, 
obtained at the Space Telescope Science Institute, which is operated by the 
Association of Universities for Research in Astronomy, Inc., under NASA 
contract NAS~5-26555.  The compilation of photometric data of the {\em Variable 
Star Network}, available at http://www.kusastro.kyoto-u.ac.jp/vsnet/ and 
managed by D.\ Nogami, was extremely useful in preparing this paper.  This work 
was supported in part by NASA grant GO-3683 to STScI and subgrants to the 
University of Washington (DWH and PS).  DWH and PS also acknowledge NASA LTSA 
grant NAGW3158.  AWS acknowledges NSF grant AST~9315280. ME acknowledges Hubble 
Fellowship grant HF-01068.01-94A.

\end{document}